# Diversity in Software Engineering: A Survey about Scientists from Underrepresented Groups

Ronnie de Souza Santos
Cape Breton University
Sydney, NS, Canada
ronnie_desouza@cbu.ca

Brody Stuart-Verner
Cape Breton University
Sydney, NS, Canada
brody_verner@cbu.ca

Cleyton V. C. de Magalhaes
CESAR School
Recife, PE, Brazil
cvcm@cesar.school

*Abstract*— Technology plays a crucial role in people's lives. However, software engineering discriminates against individuals from underrepresented groups in several ways, either through algorithms that produce biased outcomes or for the lack of diversity and inclusion in software development environments and academic courses focused on technology. This reality contradicts the history of software engineering, which is filled with outstanding scientists from underrepresented groups who changed the world with their contributions to the field. Ada Lovelace, Alan Turing, and Clarence Ellis are only some individuals who made significant breakthroughs in the area and belonged to the population that is so underrepresented in undergraduate courses and the software industry. Previous research discusses that women, LGBTQIA+ people, and non-white individuals are examples of students who often feel unwelcome and ostracized in software engineering. However, do they know about the remarkable scientists that came before them and that share background similarities with them? Can we use these scientists as role models to motivate these students to continue pursuing a career in software engineering? In this study, we present the preliminary results of a survey with 128 undergraduate students about this topic. Our findings demonstrate that students' knowledge of computer scientists from underrepresented groups is limited. This creates opportunities for investigations on fostering diversity in software engineering courses using strategies exploring computer science's history.

## I. Introduction

Technology plays a crucial role in people's lives, influencing several aspects of modern society, such as work, education, politics, and leisure. Modern society is multicultural, plural, and diverse. However, software engineering, the primary area responsible for building technology, lacks diversity in academia and industry. Software engineering discriminates, and if it does not strive to be inclusive in all its facets (i.e., education, research, and industry), the risk of harming people will be exceptionally high [1][2][3].

Computational algorithms and decision-making systems (e.g., machine learning-based algorithms) systematically discriminate against non-white communities as biased implementations prioritize the needs of some individuals and reduce access of other ethnic groups to services and opportunities [5]. Algorithms for digital filters frequently used in social media apps are designed to whiten people's skin [6].

Gender imbalance in software engineering is another problem that begins before individuals start working in the software industry [7] [8], as women are extremely underrepresented in technology courses [9]. In addition, software development environments are hostile and sexist [10]. LGBTQIA+ students are also likely to drop from these courses due to a low sense of belonging and frequent bullying [11].

The lack of diversity in software engineering is an intriguing phenomenon considering the history of this field and the renowned scientists from underrepresented groups that developed technologies that changed the world [15][17][23]. The first programmer in the world was a woman; the man who created the theory behind the modern computer was gay, and the person who developed the concepts that enable the real-time collaboration processes largely used nowadays was a Black man. These are just a few outstanding individuals among technology pioneers from underrepresented groups.

Considering that many studies discussing diversity in computer science and software engineering education have pointed out that students from underrepresented groups often feel excluded in these courses [10] [11] [12], we aim to investigate how technology groundbreakers can be used as role models to motivate students in continue pursuing a career in software engineering and consequently support the increase of diversity in software engineering. In particular, the software industry needs diversity to foster creativity and innovation, allowing it to continue developing technological solutions that reflect the reality of our society [27], and increasing diversity in undergraduate courses is the first step to effectively solving this problem.

In this paper, we present the preliminary findings obtained from a survey that was designed to assess what undergraduate students know about computer scientists from underrepresented groups. Using these preliminary findings, strategies can start to be created to explore how these scientists can be used as role models to increase diversity and inclusion in computer science and software engineering. Thus, at this initial stage, this study aimed to answer the following general research question:

*RQ. What do technology students know about the history of underrepresented groups in Software Engineering?*

From this introduction, the present study is organized as follows. Section 2 presents a background on the history of underrepresented groups in technology. Section 3 presents how we designed and conducted the survey. In section 4, we present the preliminary results obtained so far, which are discussed in section 5. In section 6, we discuss our future work. Finally, in section 7, we present our conclusions.

## II. BACKGROUND

History shows that humanity has created several mechanisms and devices to implement computing over time. Manual arithmetic, abacus, and analogic machines; they all existed long before the development of what we now recognize as the computer. The initial efforts toward developing the first computer models are from the end of the 19th century. After this period, many remarkable scientists contributed to transforming computers into the essential tool that supports our modern society [13].

Although software engineering is currently demonstrated to be a heterogenous and non-diverse field [1], mainly composed of heterosexual White men, the history of technology demonstrates that pioneers and groundbreakers scientists in the context of technology are from diverse backgrounds (i.e., cultures, regions, and education) [13]. In particular, over the decades, individuals from different underrepresented groups, such as women, Black people, LGBTQIA+ individuals, and many others, have significantly contributed to the evolution of computers and software.

Ada Lovelace and Grace Hopper are notable women in computer science [14]. The first one is recognized as being the first programmer in the world [15], while the second one was among the first modern computer programmers, and relevant contributions include the work on compilers and the fundamentals of what we know today as software testing [16]. In fact, history demonstrates that programming was initially seen as a female occupation, while hardware design was primarily performed by males [1].

The LGBTQIA+ community also participated in technological breakthroughs over the decades. Alan Turing is well-known as the *father of computer science*, as he is responsible for creating the machine that preceded modern computers and the fundamental concepts of artificial intelligence [18]. Christopher Strachey is recognized for being a pioneer in programming languages, and his studies were essential to developing the C programming language; in addition, he is also associated with the early development of video games [20]. Peter Landin developed the concepts that allow programming languages to be used in any computer (and not tied up to machines) and was a pioneer in the functional programming paradigm [21].

Clarence Ellis was the first African-American to earn a Ph.D. in computer science and an important name in computer science history as he was a pioneer in the context of computer-supported cooperative work, and his research helped in the development of tools and techniques that enable real-time collaborative editing of documents [22]. Annie J. Easley was one of the first black employees at NASA and worked on algorithms that analyzed alternative power technologies to solve energy consumption problems [23]. Similarly, Katherine Johnson and Dorothy Vaughan are two remarkable Black women who made significant contributions to technology at NASA [37]. Frank Greene was the first black cadet to make it through the U.S. Air Force and contributed to developing high-speed semiconductor computer-memory systems [23].

The history of computer science is filled with talented individuals from underrepresented groups who made enormous contributions to the development of technologies. Nowadays, these individuals are also working on innovative solutions that will continue supporting our society. Among modern scientists, we can cite Farida Bedwei, a Ghanaian software engineer with cerebral palsy [24] who is raising discussions on inclusion in software engineering; Christine M'Lot an indigenous educator working on teaching computer science and coding through music [25]; Lynne Conway, a transgender computer scientist and pioneer in conceiving new methods to simplify the design and fabrication of complex microchips [26].

## III. METHOD

Following the guidelines to conduct cross-sectional surveys in software engineering [28][29] we built a questionnaire to investigate what undergraduate students know about computer scientists from underrepresented groups who were pioneers in developing computer technologies. The study was designed as follows.

### A. Defining Objectives

Considering that computer science and software engineering undergraduate students from underrepresented groups often decide to drop these courses for the feeling of not being excluded and not belonging in these environments [10] [11] [12], the primary goal of this survey was to address their general knowledge about computer scientist from underrepresented groups and their contributions to the area. This is the first step to explore how the scientists' history can be used to motivate and engage students (e.g., through role modeling), resulting in increasing diversity awareness and promoting inclusion in software engineering education.

### B. Crafting Instruments

Before designing the questionnaire, we conducted an ad-hoc review focused on the history of computer science to identify scientists that belonged to underrepresented groups and were pioneers in developing theories, concepts, and technologies that transformed the field. After an extensive review, we selected six scientists that made breakthrough discoveries and innovative solutions that changed the way computers and software were used at the time: Ada Lovelace, Alan Turing, Christopher Strachey, Clarence Ellis, Grace Hopper, and Peter Landin. These six scientists were selected from a list of important individuals who created basic concepts and fundamentals that are taught in most computer science programs, e.g., programming, artificial intelligence, and distributed systems.

We started the questionnaire with demographic questions about the student's academic backgrounds. Next, we included an open-ended question asking students to name relevant computer scientists they know. Then, we asked them in a closed-ended question to select which scientists they know from your list (i.e., those six selected in the ad-hoc review). Following this, we added separate sections for each scientist and asked students to select what facts they knew about them (facts were both about their contributions and their life). We also asked students to point out how they learned about the scientists. We ended the questionnaire with another group of optional demographic questions about the participants' personal backgrounds.

Once all questions were defined, a pilot questionnaire was validated by two computer science professors. Following the authors' backgrounds, the questionnaire was built in Portuguese and English. Table I presents the final questionnaire in English.

TABLE I. SURVEY QUESTIONNAIRE

| Consent and Participation |
|---|
| This is an ANONYMOUS survey. Answering it is optional, but if you do be sure that no information provided by you can be related back to you. The questions you will answer are mainly about scientists from Technology and Computer Science. In addition, there are some demographic questions about you. But again, ANONYMOUS. There are no correct or incorrect answers in this survey. This survey takes less than 5 minutes. Do you agree to participate?<br>( ) Yes |
| **Course Demographics** |
| In which country is your university located?<br>What undergrad program are you enrolled in?<br>( ) Computer Science           ( ) Software Engineering<br>( ) Computer Engineering    ( ) Information Systems<br>( ) Another Technology course<br><br>What year are you in right now? |
| **Computer Scientists** |
| Name up to 3 important scientists in the history of computer/technology. |
| **Computer Scientists from Underrepresented Groups** |
| Select below all the scientists in technology that you know. Remember, there are no right or wrong answers here. We just want to know whom you have heard of before.<br>( ) Ada Lovelace           ( ) Alan Turing          ( ) Clarence Ellis<br>( ) Christopher Strachey    ( ) Grace Hopper.     ( ) Peter Landin |
| **Ada Lovelace** |
| Select below what you know about ADA LOVELACE.<br>( ) Wrote the first computer program in the world.<br>( ) Predicted that computers could do more than just crunch numbers.<br>( ) Have a programming language named after her.<br>( ) Had her contributions recognized only one century after her death.<br>( ) I don't know any facts about her. |
| **Alan Turing** |
| Select below what you know about ALAN TURING.<br>( ) Designed the basis of the modern computer.<br>( ) Created the algorithm that saved millions of lives in World War II.<br>( ) Developed the first substantial study on artificial intelligence.<br>( ) Was persecuted for being gay.<br>( ) I don't know any facts about him. |
| **Christopher Strachey** |
| Select below what you know about CHRISTOPHER STRACHEY.<br>( ) Is usually recognized as the first developer of a videogame.<br>( ) Created fundamental concepts for the C programing language.<br>( ) Developed the first research on time-shared computers.<br>( ) Was a gay man.<br>( ) I don't know any facts about him. |
| **Clarence Ellis** |
| Select below what you know about CLARENCE ELLIS.<br>( ) Was the first African-American to earn a Ph.D. in Computer Science<br>( ) Was a pioneer in Computer Supported Cooperative Work.<br>( ) Developed techniques that enable real-time collaborative editing.<br>( ) Was one of the very few African-American students in his school.<br>( ) I don't know any facts about him. |
| **Grace Hopper** |
| Select below what you know about GRACE HOPPER.<br>( ) Her team reported the first computer bug.<br>( ) Designed one of the first compilers.<br>( ) Developed the COBOL programming language.<br>( ) Was the first Woman to earn a Ph.D. in Mathematics at Yale.<br>( ) I don't know any facts about her |
| **Peter Landin** |
| Select below what you know about PETER LANDIN.<br>( ) Designed concepts that allow programming on any computer.<br>( ) Was a pioneer in the functional programming paradigm.<br>( ) Defended LGBTQIA+ rights.<br>( ) Was a bisexual man.<br>( ) I don't know any facts about him. |

In addition, each separate section about the computer scientists included the following question "*You heard about <scientist_name> through:*" and provided them with the following options: a) a course or lecture; b) an assignment or research; c) a movie or video; d) a blog or post on social media; e) I haven't heard about this person until now. The motivation behind this question was to understand how the history of computer science is being spread across the academic community of undergraduate students.

*C. Data Collection*

We followed two techniques of data collection. As the population of software engineering undergraduate students is practically impossible to determine accurately, both techniques were based on non-probability sampling, that is, sampling methods that do not employ randomness. First, we used convenience sampling and sent the questionnaire to computer science and software engineering professors. We asked them to forward the questionnaire to undergraduate students in their universities, e.g., using email lists. Second, we relied on snowballing sampling as we asked professors who answered our contact to forward our questionnaire to colleagues and collaborations.

The first round of data collection happened between November 1$^{st}$ and December 31$^{st}$, 2022. During this period, we sent our questionnaire to 65 professors from 28 countries: Argentina, Australia, Austria, Brazil, Chile, China, Colombia, Cuba, Denmark, Finlandia, Germany, Iceland, Israel, Italy, Jamaica, Japan, Netherlands, New Zealand, Norway, Portugal, Singapore, Spain, Sweden, Switzerland, Thailand, United Kingdom, Uruguay, and the United States. These professors were contacted using convenience sampling by purposefully selecting them from the program committee of conferences focused on software engineering, human aspects in software engineering, and computer science education.

As a result, twelve professors replied to our email: seven of them informing us they could share the questionnaire with students; two of them told us that they are not teaching at the undergraduate level and could not share the questionnaire; one of them said that it is against the university policy to share surveys with students; one of them requested for the ethic's board approval for the research but has not replied after receiving the requested document. The remaining 53 professors contacted did not answer the email.

With students mainly coming from one country, we decided to pause data collection to review our strategy and identify other approaches to replicate the questionnaire in other countries, possibly by searching for collaborations willing to translate the questionnaire to different languages and apply it in their region, as a family of replications [32].

*D. Data Analysis*

We applied descriptive statistics [31] to analyze the data of this survey, in particular because the data collected in this stage were mainly quantitative. Descriptive statistics allowed us to present the distribution and the frequency of students' answers.

*E. Ethics*

Following the ethics protocol approved by the first author's university, no information that could be used to identify the participants was collected in this study (e.g., name, e-mail, or university).

IV. RESULTS

At this stage, we received 128 questionnaires completed. Although we have participants from four countries, most students are from Brazil. Two factors can explain this outcome. First, one of the questionnaires was designed in

Portuguese, which facilitated the interaction with the targeted population in that country. However, we have a questionnaire written in English, but it did not receive many answers from English speakers (which the second factor can explain). Second, among the professors contacted and asked to contribute to our research by sharing the questionnaire, most of the replies to our emails came from Brazilian academics. Therefore, as for now, our sample has regional and cultural limitations. However, by analyzing our demographics, we obtained a satisfactory level of diversity regarding the participants' gender, sexual orientation, and ethnicity. Table II summarizes our sample.

TABLE II. DEMOGRAPHICS

| | Participants Profiles | |
|---|---|---|
| Country | Brazil | 112 |
| | Canada | 8 |
| | Finland | 5 |
| | US | 3 |
| Course | Computer Science | 58 |
| | Information Systems | 34 |
| | Computer Engineering | 20 |
| | Other Technology Courses | 12 |
| | Software Engineering | 4 |
| Year | 2nd Year | 37 |
| | 1st Year | 30 |
| | 4th Year | 25 |
| | 3rd Year | 21 |
| | 5th Year | 15 |
| Gender | Men | 83 |
| | Women | 39 |
| | Non-binary* | 3 |
| | Prefer not to respond | 3 |
| Sexual Orientation | Heterosexual | 85 |
| | Bisexual | 23 |
| | Gay | 6 |
| | Lesbian | 2 |
| | Pansexual | 2 |
| | Asexual | 1 |
| | No answer | 9 |
| Ethnicity | White | 86 |
| | Mixed-race | 26 |
| | Black | 9 |
| | Asian | 4 |
| | No answer | 3 |

* One non-binary participant is a transgender person

### A. Do students know computer scientists from underrepresented groups?

To answer this question, we analyzed what students spontaneously responded to when asked about renowned scientists in the history of computer science and what they answered when they selected the options from the list we provided. When students were asked to name up to three computer scientists, Alan Turing (87 mentions) and Ada Lovelace (76 mentions) were the most mentioned. In addition, the list included Grace Hopper (16 mentions) and scientists who are not considered part of an underrepresented group, such as Linus Torvalds (18 mentions), John Von Neuman (14 mentions), Charles Babbage (9 mentions), Dennis Ritchie (6 mentions), and Tim Berners-Lee (3 mentions).

Having Alan Turing and Ada Lovelace at the top of the mentions is an interesting finding because this means that, in general, students are aware of their contributions to technology, which means that facts about their personal life could be incorporated into computer science and software engineering courses to increase diversity awareness and motivate students from underrepresented groups, in these particular cases, women and LGBTQIA+ people.

Asking students to spontaneously cite relevant scientists in the history of computer science also revealed that historical aspects of this field might be overlooked in undergraduate courses, as many students had mistaken scientists and researchers for personalities and influencers from the media. At least 17 students mentioned software companies' CEOs, social media investors, and video producers (also known as internet influencers) to answer this question.

Following this, when participants were asked explicitly about the six scientists listed in our questionnaire who belonged to underrepresented groups, all of them were recognized, and the results demonstrate Alan Turing and Ada Lovelace as the most known among students, with 123 and 105 mentions, respectively, followed by Grace Hopper (78 mentions), Peter Landin (12 mentions), Clarence Ellis (3 mentions) and Christopher Strachey (2 mentions).

In summary, our findings demonstrate that students know computer scientists from underrepresented groups. Alan Turing (LGBTQIA+), Ada Lovelace, and Grace Hopper (Women) are the most recognized by them. On the other hand, professors could help students acquire more knowledge about notable scientists from other minorities, for instance, from Black and Afro-descendant scientists like Clarence Ellis.

### B. What do students know about computer scientists from underrepresented groups?

Knowing the scientists from underrepresented groups, e.g., being able to name them, does not mean that students know about their work and contributions to technology and, more importantly, their background. Learning about these scientists' life is essential for technology students from underrepresented groups because they can relate to them and consequently be motivated by them.

Table III presents the distributions of answers for each factor included in the questionnaire about the six scientists listed in our study. We observed that students are well-informed about the professional and personal backgrounds of Alan Turing and Ada Lovelace. They also demonstrated to have some knowledge about Grace Hopper. However, their knowledge of Christopher Strachey, Clarence Ellis, and Peter Landin is scarce.

We need further analysis to raise some hypotheses; however, at this point, we observed that those who knew more about Clarence Ellis were students who self-declared as being Black or Mixed-race. In addition, our findings demonstrated that those who knew about Peter Landin's personal background (e.g., being a bisexual man) were bisexual students. Further investigations will review more connections between the students' backgrounds and their knowledge of the scientists, including how they related to them.

In addition, our findings demonstrate the source of information used by students to learn about scientists from underrepresented groups. The distribution of information sources is the following: Ada Lovelace, primarily lectures and social media; Alan Turing, primarily lectures and movies; Christopher Strachey, primarily lectures and social media; Clarence Ellis, primarily social media and research; Grace Hopper, primarily lectures and research; Peter Landin, primarily lectures and social media. This result demonstrates the importance of improving the teaching of computer science history in undergraduate courses.

TABLE III. WHAT DO STUDENTS KNOW ABOUT COMPUTER SCIENTISTS FROM UNDERREPRESENTED GROUPS

| Scientist | Fact | # of Students |
|---|---|---|
| Ada Lovelace | Wrote the first computer program in the world. | 96 |
| | Had her contributions recognized only one century after her death. | 64 |
| | Have a programming language named after her. | 47 |
| | Predicted that computers could do more than just crunch numbers. | 40 |
| | I don't know any facts about her. | 24 |
| Alan Turing | Created the algorithm that saved millions of lives in World War II. | 111 |
| | Was persecuted for being gay. | 110 |
| | Designed the basis of the modern computer. | 100 |
| | Developed the first substantial study on artificial intelligence. | 58 |
| | I don't know any facts about him. | 6 |
| Christopher Strachey | I don't know any facts about him. | 117 |
| | Is usually recognized as the first developer of a videogame. | 7 |
| | Created fundamental concepts for the C programming language. | 5 |
| | Developed the first research on time-shared computers. | 3 |
| | Was a gay man. | 3 |
| Clarence Ellis | I don't know any facts about him. | 118 |
| | Was the first African-American to earn a Ph.D. in Computer Science. | 8 |
| | Was one of the very few African-American students in his school. | 4 |
| | Was a pioneer in Computer Supported Cooperative Work. | 1 |
| | Developed techniques that enable real-time collaborative editing. | 1 |
| Grace Hopper | Her team reported the first computer bug. | 57 |
| | I don't know any facts about her | 57 |
| | Developed the COBOL programming language. | 47 |
| | Designed one of the first compilers. | 36 |
| | Was the first Woman to earn a Ph.D. in Mathematics at Yale. | 24 |
| Peter Landin | I don't know any facts about him. | 118 |
| | Was a pioneer in the functional programming paradigm. | 7 |
| | Designed concepts that allow programming on any computer. | 4 |
| | Defended LGBTQIA+ rights. | 2 |
| | Was a bisexual man. | 2 |

## V. DISCUSSIONS

We focus our discussions on the use of role models as an approach to motivate undergraduate students. Role models are individuals who have their behavior admired, and often, their actions are imitated since their attitudes motivate others [33]. In healthcare, role modeling is used as a strategy to engage and support students in developing their professional skills [34]. Previous research demonstrated that teachers are used as role models by students to improve self-perception and self-discovery in the career they are pursuing [35]. The literature states that role modeling can benefit students, helping them to develop their self-esteem and motivation [36].

While researchers discuss that students from underrepresented groups are more likely to drop technology-related courses for not feeling welcomed and included in those environments, our preliminary results demonstrate that their knowledge about renowned scientists who belong to these groups is currently scarce. In addition, recent studies observed trends in the importance of representation and role models (e.g., modern scientists) to motivate and engage LGBTQIA+ students in STEM [38]. At this stage, our results cannot agree or disagree with this statement as we have not asked students specific questions about who their role models are. Our main goal was to explore how well students know historical scientists from underrepresented groups.

Our results demonstrated that students have insufficient classroom lessons about many renowned individuals from underrepresented groups who were pioneers in technologies that changed software engineering, particularly Black people and LGBTQIA+ individuals. We reinforce that educators have a vital role in changing this scenario by creating strategies to increase the visibility of underrepresented groups in computer science and software engineering. We expect that using scientists from these groups as role models (e.g., by incorporating discussions about their biographies into lectures and assignments) can foster diversity in software engineering courses by increasing inclusion and diversity awareness among students. This will be the focus of our future investigations.

## VI. FUTURE WORK

As a work in progress, this research can be improved and extended in several ways. In particular, our next step is to replicate our survey after improving the questionnaire to include: a) explicit questions about role models; b) facts about scientists from other underrepresented groups to include individuals from other countries and cultures and individuals with disabilities; c) detractors and scientist from outside of the scope of our research to increase the validity of results. In addition, we plan to conduct qualitative studies with educators to understand what strategies are currently being implemented to increase the visibility of underrepresented groups in computer science and software engineering programs. This investigation will allow comparing the perspective of students and educators about equity, diversity, and inclusion. Finally, based on the obtained findings, we aim to design strategies to support the use of role modeling in CS and SE education, similar to the successful use of this practice in other areas, such as Healthcare.

## VII. CONCLUSIONS

We surveyed 128 undergraduate students from computer science, software engineering, and other technology courses to investigate what they know about scientists from underrepresented groups who were pioneers in developing technologies that they study in class on a regular basis. This paper presented our preliminary results. In summary, students' knowledge of scientists from underrepresented groups is limited and mostly restricted to individuals with their biographies portrayed in films. We claim the need to discuss their accomplishments in lectures and academic activities to inspire and motivate students from underrepresented groups and, at the same time, foster diversity awareness among the software engineering academic community.

ACKNOWLEDGMENT

We acknowledge all the professors that supported this research by sharing the survey questionnaire across their universities and all the students who took the time to answer the questions.